\documentclass[conference]{IEEEtran}
\IEEEoverridecommandlockouts
\usepackage{cite}
\usepackage{amsmath,amssymb,amsfonts}
\usepackage{algorithmic}
\usepackage{graphicx}
\usepackage{textcomp}
\usepackage{xcolor}

\usepackage{hyperref}
\usepackage{caption}
\usepackage{subcaption}

\def\BibTeX{{\rm B\kern-.05em{\sc i\kern-.025em b}\kern-.08em
    T\kern-.1667em\lower.7ex\hbox{E}\kern-.125emX}}
\begin{document}

\title{Towards Building an End-to-End Multilingual Automatic Lyrics Transcription Model\\
}

\author{\IEEEauthorblockN{Jiawen Huang and Emmanouil Benetos\thanks{JH is a research student at the UKRI Centre for Doctoral Training in Artificial Intelligence and Music, supported jointly by UK Research and Innovation [grant number EP/S022694/1] and Queen Mary University of London. EB is supported by RAEng/Leverhulme Trust Research Fellowship LTRF2223-19-106.} }
\IEEEauthorblockA{\textit{Centre for Digital Music, Queen Mary University of London, London, UK} \\
\{jiawen.huang, emmanouil.benetos\}@qmul.ac.uk}
}

\maketitle

\begin{abstract}
Multilingual automatic lyrics transcription (ALT) is a challenging task due to the limited availability of labelled data and the challenges introduced by singing, compared to multilingual automatic speech recognition. 
Although some multilingual singing datasets have been released recently, English continues to dominate these collections.
Multilingual ALT remains underexplored due to the scale of data and annotation quality.
In this paper, we aim to create a multilingual ALT system with available datasets. Inspired by architectures that have been proven effective for English ALT, we adapt these techniques to the multilingual scenario by expanding the target vocabulary set. We then evaluate the performance of the multilingual model in comparison to its monolingual counterparts. Additionally, we explore various conditioning methods to incorporate language information into the model. We apply analysis by language and combine it with the language classification performance. Our findings reveal that the multilingual model performs consistently better than the monolingual models trained on the language subsets. Furthermore, we demonstrate that incorporating language information significantly enhances performance.

\end{abstract}

\section{Introduction}
\label{sec:intro}
Automatic lyrics transcription (ALT) is the task of recognising lyrics from singing voice. 
Access to lyrics enriches the listening experience by arousing sympathy and building a deeper connection between the music and the listeners. Moreover, lyrics transcription can benefit other music analysis tasks as well, including lyrics alignment \cite{DBLP:conf/icassp/DemirelAD21}, singing pronunciation analysis \cite{DBLP:conf/eusipco/DemirelAD21}, and cover song identification \cite{DBLP:conf/ismir/VaglioHMR21}.
While ALT shares the same input/output format and similar objectives with automatic speech recognition (ASR), it is more challenging due to larger variations in rhythm, pitch, and pronunciation \cite{DBLP:conf/dagstuhl/FujiharaG12, DBLP:conf/ismir/Kruspe14}. 
In recent years, significant progress has been achieved in ALT for English songs using end-to-end models \cite{DBLP:conf/ismir/OuGW22, DBLP:journals/taslp/GaoGL23, DBLP:conf/icassp/StollerDE19}. 
Stoller et al. \cite{DBLP:conf/icassp/StollerDE19} developed the first end-to-end lyrics alignment model using the Wave-U-Net architecture and connectionist temporal classification (CTC) loss \cite{graves2006connectionist}.  
Gao et al. enhanced performance by fine-tuning the ALT model in conjunction with a source separation frontend \cite{DBLP:journals/taslp/GaoGL23}.
Ou et al. \cite{DBLP:conf/ismir/OuGW22} leveraged wav2vec2 features and applied transfer learning techniques, resulting in a significant performance boost.

Multilingual lyrics transcription, however, remains under-explored due to the limited publicly available training and evaluation data. Although DALI v2 \cite{meseguer2020creating} is a singing dataset of moderate size with lyrics annotations, it is dominated by English songs, comprising over 80\% of the dataset.
Whisper was introduced by OpenAI \cite{DBLP:conf/icml/RadfordKXBMS23}. It is a robust ASR model trained on numerous audio-transcript pairs collected from the Internet, and the training data remains unreleased. This model has demonstrated its effectiveness in multilingual ALT \cite{DBLP:conf/ismir/ZhuoYPMLZLDFLBC23}. Building on this work, Wang et al. \cite{wang2023adapting} investigated the potential of adapting it to Mandarin Chinese ALT. A multilingual ALT dataset called MulJam was created by post-processing Whisper's output \cite{DBLP:conf/ismir/ZhuoYPMLZLDFLBC23}. 
With access to these datasets, we develop multilingual ALT models using publicly available data by combining DALI and MulJam.

Compared to English ALT, multilingual ALT faces several additional challenges: Firstly, unless explicitly specified, models must implicitly identify the underlying language of the singing to ensure that the predicted lyrics match the correct character set.
Secondly, there is a language imbalance in the datasets. Languages like English typically dominate the majority of songs, while other languages are considered low-resource in the context of ALT development. Thirdly, specific characters appear in different languages' alphabets but adhere to different pronunciation rules, adding complexity to the problem.

The multilingual ASR task shares many of the challenges mentioned above. Previous research has extensively studied and compared mono-, cross-, and multilingual models using a single model \cite{DBLP:conf/icassp/HeigoldVSNRDD13, DBLP:conf/icassp/ToshniwalSWLMWR18}. These works demonstrate that multilingual models tend to perform better than their mono-/cross-lingual counterparts, particularly in low-resource settings. Furthermore, some studies observe performance gains by incorporating language information through conditioning \cite{DBLP:conf/icassp/ToshniwalSWLMWR18} or predicting language-specific tokens \cite{DBLP:journals/corr/abs-1806-05059}. More recent research takes advantage of unlabeled data through self-supervised learning, such as wav2vec2 \cite{DBLP:conf/nips/BaevskiZMA20, DBLP:conf/interspeech/ConneauBCMA21}.

\begin{figure*}[t!]
     \centering
     \begin{subfigure}[t]{0.26\textwidth}
         \centering
         \includegraphics[width=\textwidth]{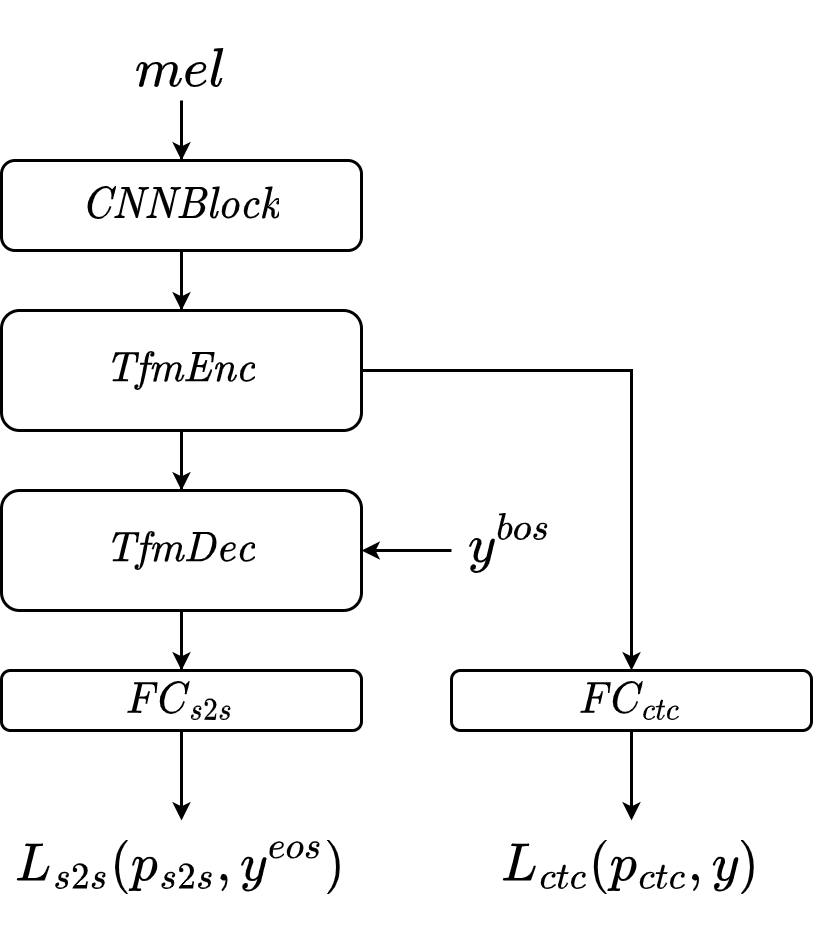}
         \caption{The multilingual and monolingual models.}
         \label{fig:multi}
     \end{subfigure}
     \hfill
     \begin{subfigure}[t]{0.26\textwidth}
         \centering
         \includegraphics[width=\textwidth]{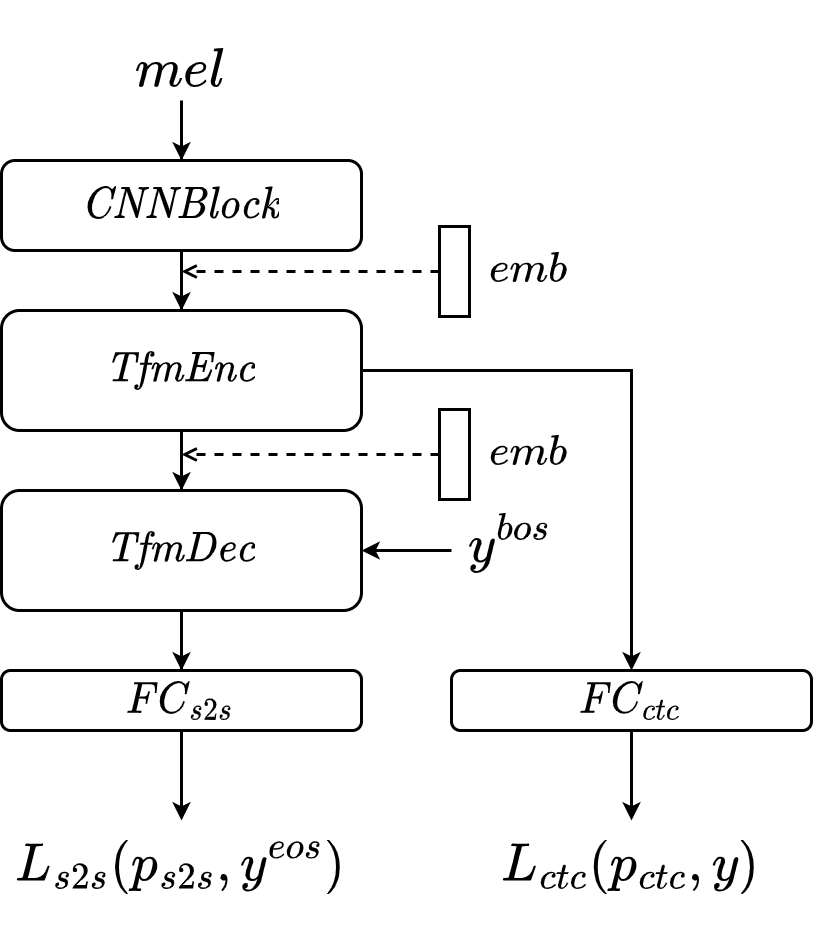}
         \caption{The language-informed model.}
         \label{fig:lang}
     \end{subfigure}
     \hfill
     \begin{subfigure}[t]{0.34\textwidth}
         \centering
         \includegraphics[width=\textwidth]{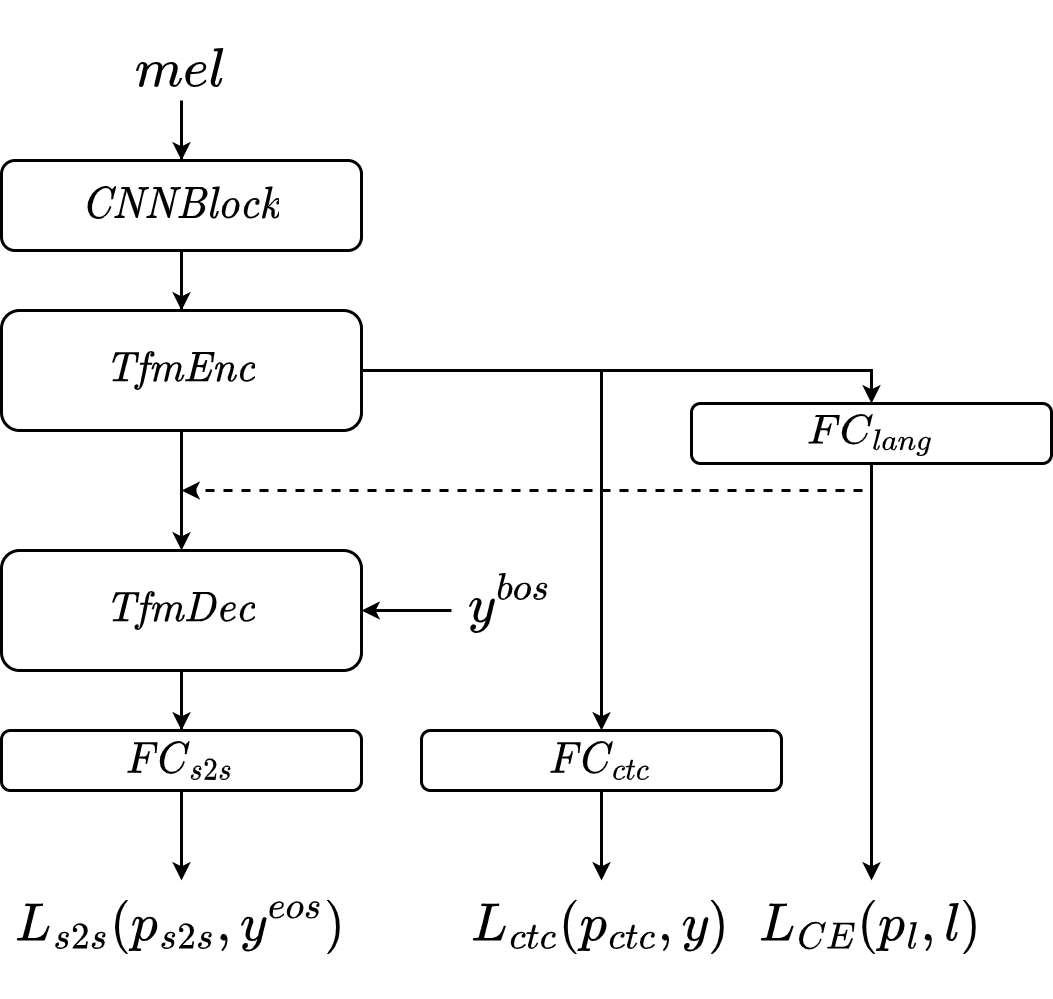}
         \caption{The language self-conditioned model.}
         \label{fig:self}
     \end{subfigure}
        \caption{Proposed model architectures at training. Our models consists of a convolutional block $\mathit{CNNBlock}$, a transformer encoder $\mathit{TfmEnc}$, a transformer decoder $\mathit{TfmDec}$, and several fully connected layers $\mathit{FC_{ctc}}$, $\mathit{FC_{s2s}}$, and $\mathit{FC_{lang}}$. The dotted lines indicate feature concatenation (with mapping for \ref{fig:self}). In \ref{fig:lang}, $emb$ denotes the language embedding.}
        \label{fig:flow}
\end{figure*}

In this work, we aim to investigate the development of multilingual ALT models using publicly accessible data, building on existing work specifically designed for English ALT. We tackle the low-resource aspects of ALT model development by training jointly on data from across a wide range of languages. Additionally, we study the impact of language information via conditioning and multi-task learning. 
Our contributions can be summarized as follows: Firstly, as one of the first attempts towards multilingual ALT, we propose a small-scale multilingual ALT model trained on publicly available datasets. Secondly, we compare the performance of multilingual models with their monolingual counterparts, revealing that training data in additional languages benefits ALT in low-resource scenarios. Thirdly, we show that language conditioning has a positive impact on performance, while the amount of improvement varies across different languages.

We acknowledge that previous research in multilingual ASR has explored similar approaches, as mentioned above. However, ALT diverges significantly due to different acoustic characteristics of the singing voice, the severe shortage of resources, and language imbalances. Therefore, fundamental assumptions need to be verified carefully. 
Our results indicate that our work will provide a solid baseline for future research, as well as an initial step in addressing the new challenges.

\section{Method}
\label{sec:method}

\subsection{Model}
Our models are built upon a similar architecture to the state-of-the-art transformer models \cite{DBLP:journals/taslp/GaoGL23}, utilising the hybrid CTC/Attention architecture \cite{8068205}. Fig.~\ref{fig:multi} illustrates this architecture at the training stage. The input is an 80-dim Mel-spectrogram computed at a sampling rate of 16kHz, with an FFT size of 400 and a hop size of 10 ms. The model consists of a convolutional block, a transformer encoder, a transformer decoder and two fully connected layers. The target dimension of the two fully connected layers is equal to the size of the target character set $N$. Let $y$ represent the target lyrics token list, $y^{bos}$ denote $y$ with a \texttt{<bos>} token added at the beginning, and $y^{eos}$ indicate $y$ with a \texttt{<eos>} token appended at the end.
During training, the Mel-spectrogram is processed through the convolutional block, before being passed to the transformer encoder and decoder. Teacher forcing is adopted for faster convergence (The model is provided with the ground truth tokens $y^{bos}$ and predicts the next ones $y^{eos}$.):
\begin{align}
    \mathit{feat} &= \mathit{CNNBlock}(mel) \\
    h &= \mathit{TfmEnc}(\mathit{feat})     \\
    o &= \mathit{TfmDec}(h, y^{bos})
\end{align}
where $h$ and $o$ are the output from the encoder and the decoder. Then $h$ and $o$ are passed to the two fully connected layers $FC_{ctc}$ and $FC_{s2s}$. These layers are responsible for generating the posteriorgrams for both the CTC branch and the sequence-to-sequence (seq2seq) branch.
\begin{align}
    p_{ctc} &= \mathit{FC_{ctc}}(h) \\
    p_{s2s} &= \mathit{FC_{s2s}}(o)
\end{align}
The loss function is a weighted sum of two components: the CTC loss for alignment-free training (computed from the CTC branch), and the Kullback-Leibler (KL) divergence loss for smooth predictions (computed from the seq2seq branch):
\begin{equation}
Loss = \alpha L_{ctc}(p_{ctc}, y) + (1-\alpha) L_{s2s}(p_{s2s}, y^{eos})
\end{equation}

\subsection{Multilingual model and monolingual models}
Let there be $M$ languages in the training set $\{\mathcal{L}_1, ..., \mathcal{L}_{M}\}$, where $\mathcal{C}_i$ represents the character set for language $\mathcal{L}_i$. In the multilingual setting, the target character set $\mathcal{C}$ is formed by taking the union of all independent character sets $\cup^{M}_{i=1}\mathcal{C}_i$.

In addition, we train individual monolingual models for each language to provide comparative analysis with multilingual models. The corresponding training and validation sets are the language-specific subsets derived from the multilingual data. 

\subsection{Language-informed models}
We condition our multilingual model with language information to study its influence. By providing the model with knowledge of the target language, it has the potential to learn language-specific features through the encoder and predict characters belonging to the target language alphabet through the decoder. 

To be more specific, an embedding is assigned to each language (Fig.~\ref{fig:lang}). During training, we explore three approaches: appending the language embedding to the input of the encoder ($\mathit{feat}$), to the input of the decoder ($h$), and to both. The three conditioned models are respectively denoted as \textbf{Enc-Cond}, \textbf{Dec-Cond}, and \textbf{EncDec-Cond}.

\subsection{Language self-conditioned model}
To gain deeper insights into the model's capability to identify the correct language, we make the language identification ability measurable by taking a multi-task learning approach (Fig.~\ref{fig:self}). The output of the encoder is averaged over time and passed to a fully connected layer $\mathit{FC_{lang}}$ to predict language ID. The predicted language probability $p_{l}$ is used as a self-conditioning vector, mapped to the embedding dimension, and appended to the input of the decoder. In this configuration, a cross-entropy loss term for language identification is added to the overall loss function, where $l$ is the language label:
\begin{equation}
\begin{split}
Loss = &\alpha L_{ctc}(p_{ctc}, y) + (1-\alpha) L_{s2s}(p_{s2s}, y^{eos}) \\
       & + \beta L_{CE}(p_{l}, l)
\end{split}
\end{equation}




\begin{table}[]
\centering
\begin{tabular}{l|cc|c|c}
\hline
   & \multicolumn{2}{c|}{Train}  & Valid   & Test    \\ 
   & DALI         & MulJam      & MulJam  & Jamendo \\ \hline \hline
English & 295444 & 85773 & 542 & 868 \\
French & 11959    & 33322  & 760 & 809 \\
Spanish & 10317    & 15146  & 566 & 881 \\
German & 26343   & 3208    & 710 & 871 \\
Italian & 9164     & 7807  & 616 & 0       \\
Russian & 0            & 1805    & 317 & 0   \\ \hline
\end{tabular}
\caption{The numbers of utterances in the training, validation, and test sets in each language.}
\label{tab:dataset}
\end{table}

\section{Experiments}

\subsection{Datasets}

The models are trained on the DALI v2 \cite{meseguer2020creating} and the MulJam \cite{DBLP:conf/ismir/ZhuoYPMLZLDFLBC23} datasets.
DALI v2 contains 7756 songs in total in more than 30 languages, among which we take the 5 languages that have more than 200 songs each: English, French, German, Spanish, and Italian. We segment the songs to line level, with paired lyrics annotations.
MulJam contains 6031 songs with line-level lyrics annotations in 6 languages: English, French, German, Spanish, Italian, and Russian. For each language, 20 songs are randomly selected for validation.

The training set is a combination of DALI-train and MulJam-train. It is important to note that the lyrics annotations in DALI do not include accented or special characters. Instead, they are converted to the Latin alphabet. Therefore, the validation and test sets for DALI v2 may not represent the real multilingual problem. We exclusively use the MulJam validation set for validation purposes.
For training and validation sets, we apply additional filtering to exclude utterances with incorrectly annotated durations, excessively long durations ($>$30s), and abnormally high character rates ($>$37.5 Hz). All utterances are source-separated by Open-Unmix \cite{DBLP:journals/jossw/StoterULM19}.

All models are evaluated on the MultiLang Jamendo dataset \cite{10096725} at line level. It consists of 80 songs in 4 languages: English, French, Spanish, and German. Line-level segments are prepared according to the line-level timestamps provided by the dataset.
Tab.~\ref{tab:dataset} shows the statistics of the data for multilingual and monolingual experiments~\footnote{The MulJam test set is not used for evaluation because 1) it is not language-balanced 2) it is too small for low-resource languages 3) lyrics annotation is provided at song-level.}.

\subsection{Model Configuration}

The convolutional block contains 3 CNN blocks with 64 channels. The first two layers have a kernel size of 5 and a stride of 2, while the last layer has both the kernel and stride set to 1. 
Positional encoding \cite{DBLP:conf/nips/VaswaniSPUJGKP17} is added to the transformer input before passing through the encoder.
The transformer encoder has 12 layers and the transformer decoder has 6 layers. Each encoder layer consists of a multi-head attention and a position-wise feed-forward layer. Each decoder layer contains the same except that the attention layer is causal. The attention dimension is set to 512, the number of heads is 4 and the position-wise feed-forward layer dimension is 2048. 

The loss weighting parameter $\alpha$ is set to 0.3, and $\beta$ is set to 0.1. The language embedding for language-informed and language self-conditioned models has a fixed size of 5 for all 6 languages. The union character set size, encompassing all 6 languages, is 91. This includes the Latin alphabet, accented and special characters, and the Cyrillic alphabet for Russian.

\begin{figure}[t]
  \centering
  \includegraphics[width =\columnwidth]{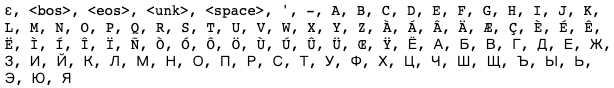}
  \caption{The multilingual vocabulary. \texttt{<bos>} and \texttt{<eos>} denote the beginning and the end of a line. \texttt{<unk>} is the unknown token. Epsilon $\varepsilon$ is included for the CTC computation.}
  \label{fig:vocab}
\end{figure}

\subsection{Training and Inference}

Our models are built upon the speechbrain \cite{speechbrain} transformer recipe for ASR~\footnote{\url{https://github.com/speechbrain/speechbrain/blob/develop/recipes/LibriSpeech/ASR/transformer/hparams/transformer.yaml} }\footnote{Our code available at: \url{https://github.com/jhuang448/MultilingualALT}}. The number of languages $M$ is 6. They are trained using the Adam optimizer \cite{DBLP:journals/corr/KingmaB14} and Noam learning rate scheduler \cite{DBLP:conf/nips/VaswaniSPUJGKP17}. The initial learning rate is 0.001 and the number of warm-up steps is 25000. The number of epochs is 50 for all models, except for the non-English monolingual ones, which have 70 epochs. The checkpoint with the lowest word error rate on the validation set is selected. 
During validation and testing, beam search is employed on the transformer decoder to select the best prediction autoregressively. The beam size is 10 at validation and 66 at testing. 
We use Word Error Rate (WER) to assess the performance of ALT models. 

\subsection{Models for comparison}
 
For our core experiments, we intentionally avoid using ALT models or feature extractors pretrained with speech data, such as wav2vec2 and Whisper, although we are aware that incorporating these could benefit the performance. This is since using speech models would introduce impact from the data distribution of the pretrained models, making it difficult to estimate if any difference is due to training on more languages' data or the knowledge acquired during pretraining.
For similar reasons, we avoid using language models. 

We still report the WER using a multilingual wav2vec2 (W2V2) variant (\texttt{large-xlsr-53} \cite{DBLP:conf/interspeech/ConneauBCMA21}) as a reference to the WER range that an adapted state-of-the-art English ALT method \cite{DBLP:conf/ismir/OuGW22} can achieve. 
The W2V2 ALT model uses a wav2vec2 feature extractor frontend, and a hybrid CTC/attention backend, similar to the proposed multilingual transformer. The CTC branch is a fully connected layer while the seq2seq branch is a one-layer recurrent neural network with attention. The attention dimension is set to 256 and the hidden size is also 256.
The target character set is the multilingual vocabulary $\mathcal{C}$. All other configurations are the same as in \cite{DBLP:conf/nips/VaswaniSPUJGKP17}.

\section{Results}

\begin{table}[]
\centering
\begin{tabular}{l|c|c|c} \hline
        & Transformer & W2V2 & Whisper\\
        & Multilingual  & XLSR-53 & large-v3\\ \hline
English & 51.45  & 42.67    & \textbf{36.80}\\
French  & 68.40  & 54.74    & \textbf{49.33} \\
Spanish & 68.02  & 45.02    & \textbf{41.15} \\ 
German  & 70.18  & 49.29    & \textbf{44.52} \\ \hline
All     & 64.31  & 47.95           & \textbf{42.95} \\ \hline
\end{tabular}
\caption{WER (\%) of the multilingual transformer, the wav2vec2-based multilingual models, and Whisper.}
\label{tab:wav2vec2}
\end{table}

\subsection{Comparison with the State-of-The-Art}
Tab.~\ref{tab:wav2vec2} lists the performance of our multilingual model, the wav2vec2-based model, and Whisper. As expected, the W2V2-based model outperforms our multilingual transformer, and Whisper performs the best due to its training with diverse data in different environments and setups.
Even when utilizing pretrained models, the WERs for non-English singing remain higher than that reported for the English monolingual model in \cite{DBLP:conf/ismir/OuGW22}, indicating greater challenges of multilingual ALT.

\begin{table}[t!]
\centering
\begin{tabular}{l|ccc} \hline
        & Monolingual & Multilingual & Self-condition   \\ \hline
English & 53.19 & 51.45  & \textbf{50.99} \\
French  & 76.06 & \textbf{68.40}  & 70.26 \\
Spanish & 79.35 & \textbf{68.02}  & 68.17 \\
German  & 82.32 & 70.18  & \textbf{67.48} \\ \hline
All     & 72.37          & 64.31  & \textbf{64.07} \\ \hline
\end{tabular}
\caption{WER (\%) of monolingual, multilingual, and language self-conditioned models.}
\label{tab:mono-multi}
\end{table}

\begin{table}[t]
\centering
\begin{tabular}{l|c|ccc} \hline
        & Multilingual & Enc-Cond    & Dec-Cond    & EncDec-Cond \\ \hline
English & 51.45  & 51.19 & \textbf{50.61} & 50.80 \\
French  & 68.40  & \textbf{65.01} & 67.33 & 65.22 \\
Spanish & 68.02  & 62.27 & 65.44 & \textbf{61.38} \\
German  & 70.18  & 63.23 & 63.95 & \textbf{62.07} \\ \hline
All     & 64.31  & 60.32 & 61.71 & \textbf{59.79} \\ \hline
\end{tabular}
\caption{WER (\%) of multilingual and language-informed models.}
\label{tab:cond}
\end{table}

\subsection{Monolingual, multilingual and language self-conditioning}

Tab.~\ref{tab:mono-multi} lists the performance of monolingual, multilingual, and the language self-conditioned models. Notably, for the monolingual models, despite having more data than Spanish, the German model yields worse WER. 
This aligns with the nature of the two languages: Spanish is more phonetic and consistent in its spelling and pronunciation than German. Additionally, German uses compounding more frequently than Spanish, leading to more variations in pronunciation and stress patterns.
This suggests that the challenge and data requirements for training ALT models vary across languages.

The multilingual model outperforms monolingual models in every language, indicating that having more training data in various languages benefits low-resource language ALT. Specifically, while the improvement for English is small ($\sim$2\%), it exceeds 7\% for all other languages. This suggests that leveraging high-resource language data (English) can be beneficial for low-resource ALT, when the target languages exhibit similarities in pronunciation and spelling rules.

Compared to the results of the multilingual model, the self-conditioned model is able to perform the additional language classification task without compromising ALT performance. It is not surprising that having the auxiliary task does not bring significant improvements to ALT, as the language class can be rather easily inferred from the predicted lyrics.

\subsection{Language-informed models and the Language Classification Accuracy}

Tab.~\ref{tab:cond} shows the performance of the multilingual and language-informed models. It can be observed that after providing the language class as input results in improved WER for all languages.
Among the three conditioning methods, \textbf{Enc-Cond} performs better than \textbf{Dec-Cond} except for English. \textbf{EncDec-Cond} gives the best overall WER, but the trend varies for each language. 
After conditioning on both encoder and decoder, there is a clear improvement for non-English languages, while the English WER remains nearly the same as that for the multilingual model. 
The French monolingual model has a better WER than Spanish and German, but this reverses after language conditioning. This indicates that with sufficient data, French ALT might be more challenging than the other two due to its complexity and frequent silent letters.

\begin{figure}[t!]
  \centering
  \includegraphics[width =0.85\columnwidth]{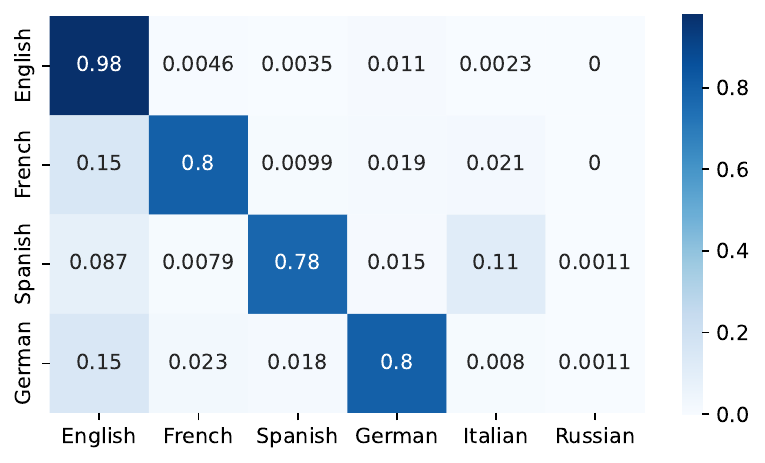}
  \caption{Confusion matrix for the self-conditioned model (\%). Languages other than English frequently get confused with English.}
  \label{fig:confusion}
\end{figure}

To gain a deeper understanding of the distinctions among languages, we examine the language classification confusion matrix of the self-conditioned model in Fig.~\ref{fig:confusion}.
As depicted, languages other than English are often misclassified as English. Additionally, Spanish singing is more commonly mistaken for Italian than English.
It is understandable that for the dominant language in the training set, the classification accuracy for English is close to 100\%, which explains why language conditioning has minimal impact on English ALT.


\section{Conclusion and future work}

In summary, our study addresses the challenges of multilingual ALT model development when training only from publicly available data, particularly in enhancing low-resource languages. We illustrate that multilingual ALT surpasses monolingual ALT for all languages, primarily due to the shared phonetic similarities among them. Our language-conditioned experiments indicate that incorporating language information enhances performance.

In this study, we take a straightforward approach by merging grapheme sets for multilingual vocabulary processing. In future research, we intend to explore more advanced strategies, including adapting target vocabularies using phoneme representations, subword units, and transliteration techniques. We believe that reducing target ambiguity is particularly critical for effective training in low-resource settings.


\bibliographystyle{IEEEtranS}
\bibliography{icassp}

\begin{thebibliography}{10}
\providecommand{\url}[1]{#1}
\csname url@samestyle\endcsname
\providecommand{\newblock}{\relax}
\providecommand{\bibinfo}[2]{#2}
\providecommand{\BIBentrySTDinterwordspacing}{\spaceskip=0pt\relax}
\providecommand{\BIBentryALTinterwordstretchfactor}{4}
\providecommand{\BIBentryALTinterwordspacing}{\spaceskip=\fontdimen2\font plus
\BIBentryALTinterwordstretchfactor\fontdimen3\font minus \fontdimen4\font\relax}
\providecommand{\BIBforeignlanguage}[2]{{%
\expandafter\ifx\csname l@#1\endcsname\relax
\typeout{** WARNING: IEEEtranS.bst: No hyphenation pattern has been}%
\typeout{** loaded for the language `#1'. Using the pattern for}%
\typeout{** the default language instead.}%
\else
\language=\csname l@#1\endcsname
\fi
#2}}
\providecommand{\BIBdecl}{\relax}
\BIBdecl

\bibitem{DBLP:conf/nips/BaevskiZMA20}
A.~Baevski, Y.~Zhou, A.~Mohamed, and M.~Auli, ``wav2vec 2.0: {A} framework for self-supervised learning of speech representations,'' in \emph{Advances in Neural Information Processing Systems 33: Annual Conference on Neural Information Processing Systems 2020, NeurIPS 2020, December 6-12, 2020, virtual}, 2020.

\bibitem{DBLP:conf/interspeech/ConneauBCMA21}
A.~Conneau, A.~Baevski, R.~Collobert, A.~Mohamed, and M.~Auli, ``Unsupervised cross-lingual representation learning for speech recognition,'' in \emph{Interspeech 2021, 22nd Annual Conference of the International Speech Communication Association, Brno, Czechia, 30 August - 3 September 2021}.\hskip 1em plus 0.5em minus 0.4em\relax {ISCA}, 2021, pp. 2426--2430.

\bibitem{DBLP:conf/eusipco/DemirelAD21}
E.~Demirel, S.~Ahlb{\"{a}}ck, and S.~Dixon, ``Computational pronunciation analysis in sung utterances,'' in \emph{29th European Signal Processing Conference, {EUSIPCO} 2021, Dublin, Ireland, August 23-27, 2021}.\hskip 1em plus 0.5em minus 0.4em\relax {IEEE}, 2021, pp. 186--190.

\bibitem{DBLP:conf/icassp/DemirelAD21}
E.~Demirel, S.~Ahlb{\"a}ck, and S.~Dixon, ``Low resource audio-to-lyrics alignment from polyphonic music recordings,'' in \emph{{IEEE} International Conference on Acoustics, Speech and Signal Processing, {ICASSP} 2021, Toronto, ON, Canada, June 6-11, 2021}.\hskip 1em plus 0.5em minus 0.4em\relax {IEEE}, 2021, pp. 586--590.

\bibitem{10096725}
S.~Durand, D.~Stoller, and S.~Ewert, ``Contrastive learning-based audio to lyrics alignment for multiple languages,'' in \emph{ICASSP 2023 - 2023 IEEE International Conference on Acoustics, Speech and Signal Processing (ICASSP)}, 2023, pp. 1--5.

\bibitem{DBLP:conf/dagstuhl/FujiharaG12}
H.~Fujihara and M.~Goto, ``Lyrics-to-audio alignment and its application,'' in \emph{Multimodal Music Processing}, ser. Dagstuhl Follow-Ups.\hskip 1em plus 0.5em minus 0.4em\relax Schloss Dagstuhl - Leibniz-Zentrum fuer Informatik, Germany, 2012, vol.~3, pp. 23--36.

\bibitem{DBLP:journals/taslp/GaoGL23}
X.~Gao, C.~Gupta, and H.~Li, ``Polyscriber: Integrated fine-tuning of extractor and lyrics transcriber for polyphonic music,'' \emph{{IEEE} {ACM} Trans. Audio Speech Lang. Process.}, vol.~31, pp. 1968--1981, 2023.

\bibitem{graves2006connectionist}
A.~Graves, S.~Fern{\'{a}}ndez, F.~J. Gomez, and J.~Schmidhuber, ``Connectionist temporal classification: labelling unsegmented sequence data with recurrent neural networks,'' in \emph{Proc. ICML}, vol. 148.\hskip 1em plus 0.5em minus 0.4em\relax {ACM}, 2006, pp. 369--376.

\bibitem{DBLP:conf/icassp/HeigoldVSNRDD13}
G.~Heigold, V.~Vanhoucke, A.~W. Senior, P.~Nguyen, M.~Ranzato, M.~Devin, and J.~Dean, ``Multilingual acoustic models using distributed deep neural networks,'' in \emph{{IEEE} International Conference on Acoustics, Speech and Signal Processing, {ICASSP} 2013, Vancouver, BC, Canada, May 26-31, 2013}.\hskip 1em plus 0.5em minus 0.4em\relax {IEEE}, 2013, pp. 8619--8623.

\bibitem{DBLP:journals/corr/KingmaB14}
D.~P. Kingma and J.~Ba, ``Adam: {A} method for stochastic optimization,'' in \emph{3rd International Conference on Learning Representations, {ICLR} 2015, San Diego, CA, USA, May 7-9, 2015, Conference Track Proceedings}, 2015.

\bibitem{DBLP:conf/ismir/Kruspe14}
A.~M. Kruspe, ``Keyword spotting in a-capella singing,'' in \emph{Proceedings of the 15th International Society for Music Information Retrieval Conference, {ISMIR}, Taipei, Taiwan, October 27-31}, 2014, pp. 271--276.

\bibitem{meseguer2020creating}
G.~Meseguer-Brocal, A.~Cohen-Hadria, and G.~Peeters, ``Creating dali, a large dataset of synchronized audio, lyrics, and notes,'' \emph{Transactions of the International Society for Music Information Retrieval}, vol.~3, no.~1, 2020.

\bibitem{DBLP:conf/ismir/OuGW22}
L.~Ou, X.~Gu, and Y.~Wang, ``Transfer learning of wav2vec 2.0 for automatic lyric transcription,'' in \emph{Proceedings of the 23rd International Society for Music Information Retrieval Conference, {ISMIR} 2022, Bengaluru, India, December 4-8, 2022}, 2022, pp. 891--899.

\bibitem{DBLP:conf/icml/RadfordKXBMS23}
A.~Radford, J.~W. Kim, T.~Xu, G.~Brockman, C.~McLeavey, and I.~Sutskever, ``Robust speech recognition via large-scale weak supervision,'' in \emph{International Conference on Machine Learning, {ICML} 2023, 23-29 July 2023, Honolulu, Hawaii, {USA}}, ser. Proceedings of Machine Learning Research, vol. 202.\hskip 1em plus 0.5em minus 0.4em\relax {PMLR}, 2023, pp. 28\,492--28\,518.

\bibitem{speechbrain}
M.~Ravanelli, T.~Parcollet, P.~Plantinga, A.~Rouhe, S.~Cornell, L.~Lugosch, C.~Subakan, N.~Dawalatabad, A.~Heba, J.~Zhong, J.-C. Chou, S.-L. Yeh, S.-W. Fu, C.-F. Liao, E.~Rastorgueva, F.~Grondin, W.~Aris, H.~Na, Y.~Gao, R.~D. Mori, and Y.~Bengio, ``{SpeechBrain}: A general-purpose speech toolkit,'' 2021, arXiv:2106.04624.

\bibitem{DBLP:conf/icassp/StollerDE19}
D.~Stoller, S.~Durand, and S.~Ewert, ``End-to-end lyrics alignment for polyphonic music using an audio-to-character recognition model,'' in \emph{{IEEE} International Conference on Acoustics, Speech and Signal Processing, {ICASSP} 2019, Brighton, United Kingdom, May 12-17, 2019}.\hskip 1em plus 0.5em minus 0.4em\relax {IEEE}, 2019, pp. 181--185.

\bibitem{DBLP:journals/jossw/StoterULM19}
F.~St{\"{o}}ter, S.~Uhlich, A.~Liutkus, and Y.~Mitsufuji, ``Open-unmix - {A} reference implementation for music source separation,'' \emph{J. Open Source Softw.}, vol.~4, no.~41, p. 1667, 2019.

\bibitem{DBLP:conf/icassp/ToshniwalSWLMWR18}
S.~Toshniwal, T.~N. Sainath, R.~J. Weiss, B.~Li, P.~J. Moreno, E.~Weinstein, and K.~Rao, ``Multilingual speech recognition with a single end-to-end model,'' in \emph{2018 {IEEE} International Conference on Acoustics, Speech and Signal Processing, {ICASSP} 2018, Calgary, AB, Canada, April 15-20, 2018}.\hskip 1em plus 0.5em minus 0.4em\relax {IEEE}, 2018, pp. 4904--4908.

\bibitem{DBLP:conf/ismir/VaglioHMR21}
A.~Vaglio, R.~Hennequin, M.~Moussallam, and G.~Richard, ``The words remain the same: Cover detection with lyrics transcription,'' in \emph{Proceedings of the 22nd International Society for Music Information Retrieval Conference, {ISMIR} 2021, Online, November 7-12, 2021}, 2021, pp. 714--721.

\bibitem{DBLP:conf/nips/VaswaniSPUJGKP17}
A.~Vaswani, N.~Shazeer, N.~Parmar, J.~Uszkoreit, L.~Jones, A.~N. Gomez, L.~Kaiser, and I.~Polosukhin, ``Attention is all you need,'' in \emph{Advances in Neural Information Processing Systems 30: Annual Conference on Neural Information Processing Systems 2017, December 4-9, 2017, Long Beach, CA, {USA}}, 2017, pp. 5998--6008.

\bibitem{wang2023adapting}
J.-Y. Wang, C.-I. Leong, Y.-C. Lin, L.~Su, and J.-S.~R. Jang, ``Adapting pretrained speech model for mandarin lyrics transcription and alignment,'' in \emph{2023 IEEE Automatic Speech Recognition and Understanding Workshop (ASRU)}.\hskip 1em plus 0.5em minus 0.4em\relax IEEE, 2023, pp. 1--8.

\bibitem{8068205}
S.~Watanabe, T.~Hori, S.~Kim, J.~R. Hershey, and T.~Hayashi, ``Hybrid ctc/attention architecture for end-to-end speech recognition,'' \emph{IEEE Journal of Selected Topics in Signal Processing}, vol.~11, no.~8, pp. 1240--1253, 2017.

\bibitem{DBLP:journals/corr/abs-1806-05059}
S.~Zhou, S.~Xu, and B.~Xu, ``Multilingual end-to-end speech recognition with {A} single transformer on low-resource languages,'' \emph{ArXiv}, vol. abs/1806.05059, 2018.

\bibitem{DBLP:conf/ismir/ZhuoYPMLZLDFLBC23}
L.~Zhuo, R.~Yuan, J.~Pan, Y.~Ma, Y.~Li, G.~Zhang, S.~Liu, R.~B. Dannenberg, J.~Fu, C.~Lin, E.~Benetos, W.~Chen, W.~Xue, and Y.~Guo, ``Lyricwhiz: Robust multilingual zero-shot lyrics transcription by whispering to chatgpt,'' in \emph{Proceedings of the 24th International Society for Music Information Retrieval Conference, {ISMIR} 2023, Milan, Italy, November 5-9, 2023}, 2023, pp. 343--351.

\end{thebibliography}

\end{document}